# A GA based approach for task scheduling in multi-cloud environment

Tripti Tanaya Tejaswi, Md Azharuddin, P. K. Jana, *IEEE Senior member*
Department of Computer Science and Engineering
Indian School of Mines Dhanbad-826004, India
tripti.tejaswi @gmail.com, azhar_ism@yahoo.in, prasantajana@yahoo.com

**ABSTRACT**

In multi-cloud environment, task scheduling has attracted a lot of attention due to NP-Complete nature of the problem. Moreover, it is very challenging due to heterogeneity of the cloud resources with varying capacities and functionalities. Therefore, minimizing the makespan for task scheduling is a challenging issue. In this paper, we propose a genetic algorithm (GA) based approach for solving task scheduling problem. The algorithm is described with innovative idea of fitness function derivation and mutation. The proposed algorithm is exposed to rigorous testing using various benchmark datasets and its performance is evaluated in terms of total makespan.

## 1. INTRODUCTION

Cloud computing is a modern technology that minimizes the complexity of IT sector by providing the economical sharing of on-demand and automatically arranged virtual infrastructure. Significantly, the computing is a basic utility like water and electricity [1]. Cloud services grant man-to-man and trades to utilize software and hardware that commanded by another party at remote areas in the form of infrastructure (IaaS), platform (PaaS), and software (SaaS) as a service [2], [3]. Cloud computing depends upon distribution of resources to achieve coherence and scalability, similar to other utility over a network. It is a kind of computing which must cater to the dynamism, abstraction and resource sharing needs etc. In IaaS, resource sharing is the main thing of computation, where objective of cloud providers is to maximize utilization of resources. Here, virtual machines (VMs) are considered as resources [2], [4]. In cloud computing, there are many applications succumbed to the datacenter to secure some events on the basis of payment vector [2], [3]. To satisfy customer requirement in minimum time [5], VMs are allotted to the applications that contain group of tasks which are dependent to each other. This infers the task scheduling problem in the environment of multi-cloud system and thus, diverse effort [6], [7], [8] have been contrived to encounter a near optimal solution. However the problem with dependent tasks applications is very vital and challenging objective.

An application is assemblage of miscellaneous tasks which is portrayed by a directed acyclic graph (DAG) [2]. In DAG some tasks are independent and some are dependent on other tasks. Every application is concluded by achieving an optimal mapping and scheduling of interdependent and independent tasks in multifarious clouds. In this way, linked tasks are committed in appropriate consecutive manner prettified by precedence among the tasks [2], [9]. It is noteworthy that cloud computing with $t$ tasks and $c$ clouds, the number of possible allocations is $c^t$. Thus, the complexity to schedule the tasks on the clouds is very large in brute force approach and obtaining minimum execution time (makespan) is NP-complete problem [8], [10], [11].

Therefore, numerous surveys [6-8] have been contrived to retrieve optimal result for task scheduling. Most of these endeavors have been accomplished by independent tasks that is executed concurrently by multiple VMs, while it needs more scrutiny when tasks are dependent [12]. The path of tasks is unsure also the direction of execution and flow of spans are not certain due to dependent tasks [13]. In the field of artificial intelligence, a genetic algorithm (GA) is a metaheuristic approach that imitates of natural selection [13] used for solving optimization problem. Therefore, it can be used in scheduling problem in order to get optimum solution.

In this paper, we propose GA based task scheduling algorithm for non-preemptive dependent tasks. We also present an efficient chromosome encoding scheme and wisely design the mutation scheme to maximize the resource utilization. This is also complimented with derivation of fitness function. Finally, we execute the simulation of proposed algorithm and depicts various results to show the effectiveness.

The paper is organized as follows: The section 2 points some related researches for task scheduling. Section 3 represents the system model of cloud. The problem statement and the introduction of genetic algorithm are presented in section 4 and 5 respectively. In section 6, we present our proposed algorithm. The section 7 expresses our simulation results and conclusion is given in section 8.

## 2. RELATED RESEARCHES

A number of task scheduling algorithms [2], [6], [8], [14] have been proposed for independent tasks in heterogeneous cloud system. Sanjay et al. [2] have described the minimum completion cloud scheduling (MCC) algorithm to observe total makespan for all appointed independent tasks to VMs. However, the authors have not considered dependency among the tasks. Cloud list scheduling (CLS) and cloud Min-Min scheduling (CMMS) algorithms have delivered by Li et al. [6] in IaaS domain and both these algorithms have deliberated on the basis of on-demand in heterogeneous multi-cloud arrangement. However, these algorithms have not analyzed their performance with benchmark dataset. Braun et al. [11] demonstrated min-min heuristic approach for scheduling separate parallel tasks in independent surroundings. This algorithm works one by one in the queue until all tasks are deal out.

Kwok and Ahmad et al. [14] illustrated a number of static scheduling for DAG to complete the tasks for enhanced results in completion time. Generally these algorithms are for parallel tasks and the time during connection are not considered. Geoffrey et al. [15] applied genetic algorithm based task scheduling in interdependent environment. But, the authors have done the crossover to maintain the aligning of dependent tasks and chromosomes are updated according to dependency. However, it changes the order of tasks.

Although many algorithms have been developed for task scheduling algorithms, but in this paper, we propose an algorithm to schedule dependent and divisible tasks adapting to different computations on different clouds using genetic algorithm. It preserves dependency among the tasks and improve the utility of clouds in the mutation part of genetic algorithm.

### 3. SYSTEM MODEL OF CLOUD

Cloud is a group of virtualized computer resources. It can provide the users a diverse range of tasks in the form of batch style backend and client facing applications by IaaS, PaaS and SaaS model shown in Figure 1. The requirement may be in terms of number of CPUs, chunk of memory and internet bandwidth. In IaaS model, numeral data centers [6] are handed over in terms of software and hardware according to demand by third party remotely. In data centers, VMs are deployed for the arrangement of resource computation.

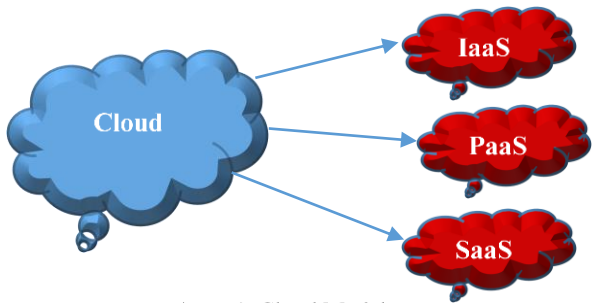

Figure 1. Cloud Models

Virtualization of server resources have empowered for both, user and provider in the sense of cost effective and better utilization. As per incoming demands, a number of VMs can begin and terminate over a single machine [3], thereby furnishing the elasticity of designing different allotment of resources on the single existing machine according to the necessity of applications. After obtaining an application in centralized organization in federated cloud model, a super computer arranges the tasks among VMs in parallel manner [2], [6]. In cloud, tasks can be completed on according to dependency and maintaining the critical section, mutual exclusion and progress.

### 4. PROBLEM STATEMENT

In this paper, we have considered that we have given an ETC (expected time to compute) matrix and a set of applications in which each application consists a set of tasks having precedence amongst them. The problem is to allocate all the tasks to the set of clouds, so that the makespan (total execution time) is minimized. Suppose a group of $p$ applications $A = \{A_1, A_2, A_3,\ldots , A_p\}$ and $q$ number of clouds $C = \{C_1, C_2, C_3,\ldots, C_q\}$ are given to complete all tasks of demanded applications. Every application $A_k$ consists a set of $r$ tasks $T = \{T_1, T_2, T_3,\ldots, T_r\}$ where $k = \{1, 2,\ldots, p\}$ and expressed by directed acyclic graph (DAG). The DAG $D = (V, E)$ represents the interdependency among tasks, where $V$ is the set of vertices pictured as tasks and $E$ is the set of edges represents precedence among tasks. An edge between two vertices $V_i$ and $V_j$ is represented by $E_{ij}$ ($E_{ij} \in E$) and ($V_i \to V_j$) denotes $V_i$ should be executed before $V_j$.

Suppose, there are a total bunch of $n$ tasks $t = \{t_1, t_2, \ldots , t_n\}$ for execution of $p$ applications $A = \{A_1, A_2, \ldots , A_p\}$. Then, the size of dependency matrix (*size_dep_mat*) for $p$ applications is represented by:

$$size\_dep\_mat = \left[\sum_{i=1}^{n} t_i\right]^2 \qquad (4.1)$$

where $n$ is the total number of tasks.

In the ETC matrix, different task consumed different time provided by different clouds. The objective is to allocate all the tasks over clouds in this fashion so that the gross computing time is minimal.

### 5. PROPOSED ALGORITHM

The basic idea of the proposed GA based algorithm is as follows. Initially, we generate $l$ number of chromosomes (population size) in the form of 1-$d$ arrays with size equal to the number of tasks. Here, we represents the assignment of a $j^{th}$ task in the $i^{th}$ chromosome to the $k^{th}$ cloud by pop[$i$][$j$] = $k$, where $1 \leq i \leq l$, $1 \leq j \leq$ number of task and number of clouds $k << j$. Now with the help of ETC matrix, we calculate the makespan of each chromosome which is the fitness function for the proposed GA based algorithm. The various steps of the proposed GA based algorithm are described below with the help of a suitable example.

Consider three distinct applications as shown in Figure 2, which are arrived at same time. Each application has a group of tasks and dependency is exhibited by the edge. Note that there are 14 tasks $T = \{A, B, C,…, N\}$ in the three distinct applications which need to be allocated to 4 clouds $C = \{C_1, z_2, C_3, C_4\}$. The execution time of the tasks on various clouds is given by an ETC matrix as shown in the Table 1. The dependency among the tasks is represented by an adjacency matrix as shown in Table 2.

## 5.1 Initial Population

The initial population is the set of randomly generated chromosomes and each chromosome denotes a solution to the task scheduling problem. Here, we show how to represent a chromosome for the example in Figure 2. As the total number of tasks is 14, the length of chromosome is also kept 14. Each task can be allocated amongst 4 clouds at random and a task is assigned to one and only one cloud. As an illustration, a chromosome representation is shown in Figure 3.

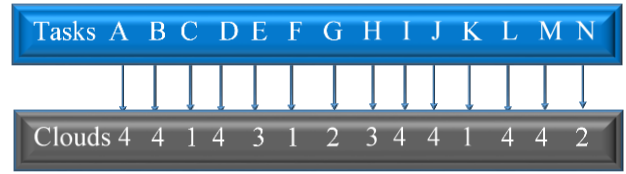

**Figure 3. Chromosome representation**

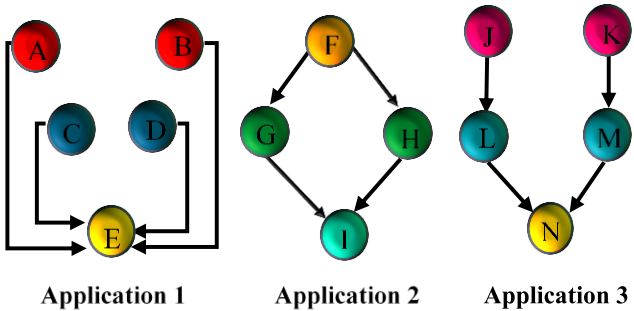

**Figure 2** The DAG representation of three application

**Table 1: An ETC matrix along with 14 tasks and 4 clouds**

|         | A  | B | C | D  | E | F | G  | H | I | J | K | L  | M | N |
|---------|----|---|---|----|---|---|----|---|---|---|---|----|---|---|
| CLOUD 1 | 6  | 7 | 8 | 10 | 4 | 5 | 6  | 7 | 3 | 7 | 9 | 4  | 5 | 5 |
| CLOUD 2 | 10 | 9 | 8 | 7  | 8 | 5 | 4  | 6 | 3 | 4 | 4 | 6  | 6 | 8 |
| CLOUD 3 | 3  | 4 | 5 | 3  | 4 | 5 | 10 | 8 | 9 | 9 | 8 | 7  | 6 | 5 |
| CLOUD 4 | 2  | 3 | 4 | 5  | 6 | 7 | 8  | 9 | 9 | 8 | 7 | 10 | 3 | 4 |

**Table 2: The dependency matrix for 14 tasks**

|   | A | B | C | D | E | F | G | H | I | J | K | L | M | N |
|---|---|---|---|---|---|---|---|---|---|---|---|---|---|---|
| A | 0 | 0 | 0 | 0 | 0 | 0 | 0 | 0 | 0 | 0 | 0 | 0 | 0 | 0 |
| B | 0 | 0 | 0 | 0 | 0 | 0 | 0 | 0 | 0 | 0 | 0 | 0 | 0 | 0 |
| C | 0 | 0 | 0 | 0 | 0 | 0 | 0 | 0 | 0 | 0 | 0 | 0 | 0 | 0 |
| D | 0 | 0 | 0 | 0 | 0 | 0 | 0 | 0 | 0 | 0 | 0 | 0 | 0 | 0 |
| E | 1 | 1 | 1 | 1 | 0 | 0 | 0 | 0 | 0 | 0 | 0 | 0 | 0 | 0 |
| F | 0 | 0 | 0 | 0 | 0 | 0 | 0 | 0 | 0 | 0 | 0 | 0 | 0 | 0 |
| G | 0 | 0 | 0 | 0 | 0 | 1 | 0 | 0 | 0 | 0 | 0 | 0 | 0 | 0 |
| H | 0 | 0 | 0 | 0 | 0 | 1 | 0 | 0 | 0 | 0 | 0 | 0 | 0 | 0 |
| I | 0 | 0 | 0 | 0 | 0 | 0 | 1 | 1 | 0 | 0 | 0 | 0 | 0 | 0 |
| J | 0 | 0 | 0 | 0 | 0 | 0 | 0 | 0 | 0 | 0 | 0 | 0 | 0 | 0 |
| K | 0 | 0 | 0 | 0 | 0 | 0 | 0 | 0 | 0 | 1 | 0 | 0 | 0 | 0 |
| L | 0 | 0 | 0 | 0 | 0 | 0 | 0 | 0 | 0 | 0 | 1 | 0 | 0 | 0 |
| M | 0 | 0 | 0 | 0 | 0 | 0 | 0 | 0 | 0 | 0 | 0 | 1 | 1 | 0 |
| N | 0 | 0 | 0 | 0 | 0 | 0 | 0 | 0 | 0 | 0 | 0 | 1 | 1 | 0 |

## 5.2 Derivation of Fitness Function

The fitness function is derived as follows. Since our basic objective is to minimize the makespan for the dependent tasks, we use makespan as the fitness function for this problem which is calculated as follows. As the completion time for a dependent task is the summation of waiting and execution time, thus before calculating the makespan for a chromosome we need to calculate the completion time for each task. It is calculated recursively as follows. Let, $TC_{i,j}$ denotes the execution time of task $T_i$ on cloud $C_j$. If the task $T_i$ depends on the task $T_k$ (symbolically, $T_k \rightarrow T_i$), then $T_i$ cannot start its execution before the completion of $T_k$. However, $T_k$ must also have to wait for the completion of its parent task $T_p$ (if any). Note that, the task $T_i$ may depend on multiple parent tasks, e.g., the task I in Figure 3 depends on the tasks G and H, in such situation the waiting time for $T_i$ is the maximum of sum of waiting and execution time of the parent tasks. Thus, the waiting time for the task $T_i$ denoted by $w(T_i)$ is calculated as follows:

$$w(T_i) = \begin{cases} 0 & \text{if } T_i \text{ is independent} \\ w(T_k) + TC_{k,j} & \text{such that } T_k \rightarrow T_i, \\ & \forall T_i, T_k \in T, \; C_j \in C \end{cases} \quad (6.1)$$

Therefore, the completion time for each task $T_i$ denoted by $\sigma(T_i)$ in the chromosome is given as follows.

$$\sigma(T_i) = w(T_i) + TC_{i,j} \quad (6.2)$$

where, $C_j \in C$ is the assigned cloud for the task $T_i$ for the chromosome. Therefore, the makespan is the sum of completion time of all the tasks in the chromosome. Therefore,

$$Makespan = \sum_{i=1}^{n} \sigma(T_i) \quad (6.3)$$

Thus, the fitness function of the proposed algorithm can be expressed as

$$Fitness\ function = Minimize\ Makespan. \quad (6.4)$$

## 5.3 Selection

We select some efficient chromosomes from the initial population for further enhancement of the chromosomes to produce better solutions with the help of crossover and mutation operation. There are many selection methods e.g., tournament selection, rank selection and roulette wheel selection etc. However, we use roulette wheel selection in order to select the best chromosomes.

## 5.4 Crossover

In crossover, some part of information of two parent chromosomes are exchanged to produce two better child chromosomes which is shown in Figure 4. To do this, we consider 1-point crossover which is taken in random manner.

## 5.5 Mutation

The quality of child chromosomes are further enhanced in the mutation operation by minimizing makespan and maximizing resource utilization. In this paper, we perform the mutation on the selected gene value rather than performing random mutation which is done in traditional GA. Here, first we find out the busiest cloud and then select a task $T_i$ randomly from it.

Now, the selected task $T_i$ may be independent or dependent. If the selected task is independent then we remove $T_i$ from the cloud and assigns it to the least utilized cloud. In case of dependent task, we also do the same as independent task. However for the dependent task, we also assign all the tasks depend on $T_i$ directly or indirectly to the same cloud i.e., the cloud where $T_i$ is assigned. The Figure 5 and Figure 6 show the chromosome representation before and after mutation for independent task.

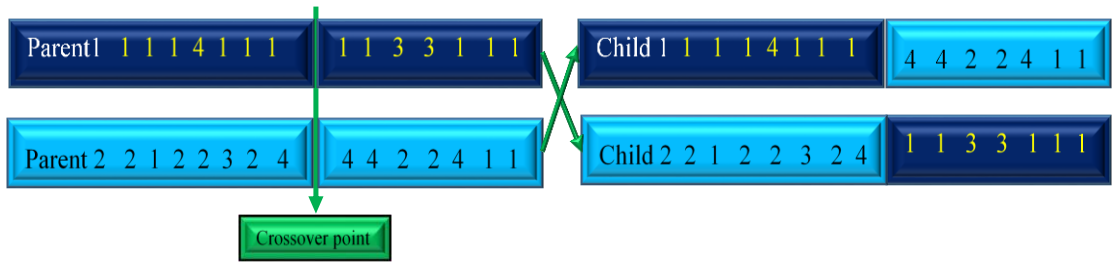

**Figure 4. Crossover**

## 6. SIMULATION RESULTS

We tested the proposed algorithm through simulation run with some benchmark. The simulations were done with the help of C++ programming language and MATLAB R2013a on an Intel Core i3 processor with 2.20 GHz CPU and 4 GB RAM running on the platform Microsoft Windows 8.1.

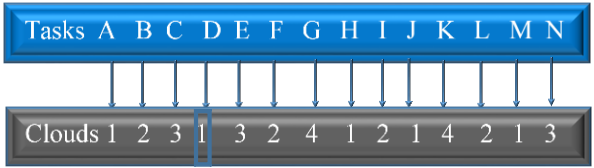

Figure 5. Before Mutation

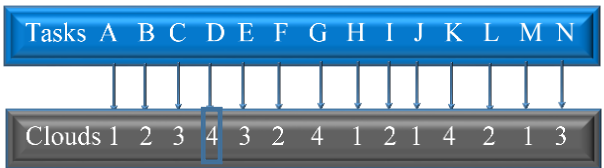

Figure 6. After Mutation

### 6.1 Simulation Results on Benchmark Dataset

For the comparison purpose, we assumed that 20 and 30 applications for each of the benchmark datasets of sizes 512× 16 and 1024 ×32 created by Braun et al. [11]. Expected time to compute for all the tasks of benchmark data sets is considered in millisecond (ms). In addition to this, we also considered a dependency matrix size of ($a \times a$), where $a$ represents the total number of tasks, which exhibit the dependency among tasks. The universal format of dataset instances is reprsented as $u\_x\_vvww$, where $u$ reprsents uniform distribution to form the instances, $x$ represents verity of consistency such as consistent($c$), inconsistent($i$) and semiconsistent($s$), $vv$ reveals task heterogeneity that is either high ($hi$) or low($lo$) and ww represents machine heterogeneity that is high and low. Hence 12 intances that is ($u\_c\_hihi$, $u\_c\_hilo$, $u\_c\_lohi$, $u\_c\_lolo$, $u\_i\_hihi$, $u\_i\_hilo$, $u\_i\_lohi$, $u\_i\_lolo$, $u\_s\_hihi$, $u\_s\_hilo$, $u\_s\_lohi$, $u\_s\_lolo$) [2]. The dependency matrix has been taken for 20 and 30 applications with the help of adjacency matrix. The makespan (ms) of the datasets $512 \times 16$ and $1024 \times 32$ compared for 20 and 30 applications for dependent task environment is shown in Table 3. This is obvious to note that the proposed algorithm performs best for the instance $u\_i\_lolo$ and moderately for the instances $u\_c\_lohi$ and $u\_s\_lohi$ for both of the applications and the data sets. We have also shown the makespan for the above datasets in the form of bar chart in Figure 7 for 20 applications which also indicate the same fact.

Table 3: Makespan for 512× 16 and 1024 ×32 dataset for 20 and 30 applications

| Instance | Cloud Makespan | | | |
|---|---|---|---|---|
| | 512×16(20 Appl.) | 512*16(30 Appl.) | 1024*32(20 Appl.) | 1024*32(30 Appl.) |
| u_c_hihi | 51769464.21 | 45738512.69 | 230547053.5 | 250200111.8 |
| u_c_hilo | 523937.961 | 466702.8287 | 24391252.86 | 20296357.06 |
| u_c_lolo | 17980.60393 | 16387.84265 | 2563.6 | 2283.39 |
| u_c_lohi | 1655205.53 | 1613689.219 | 24362.52 | 22807.66 |
| u_i_hihi | 45350660.72 | 41168665.13 | 219098889.8 | 180804001.2 |
| u_i_lohi | 1431057.056 | 1261567.662 | 21373.33 | 16909.65 |
| u_i_hilo | 485206.1876 | 410295.7933 | 20110770.02 | 17121479.97 |
| u_i_lolo | 15521.7232 | 15708.7488 | 2288.67 | 1743.17 |
| u_s_hihi | 47602798.46 | 43741044 | 235605168.5 | 182364945.8 |
| u_s_hilo | 477202.8942 | 451603.9588 | 24377455.16 | 20375404.55 |
| u_s_lolo | 17238.50465 | 16252.38443 | 2370.65 | 1918.45 |
| u_s_lohi | 1386096.756 | 1348281.766 | 23092.17 | 19955.34 |

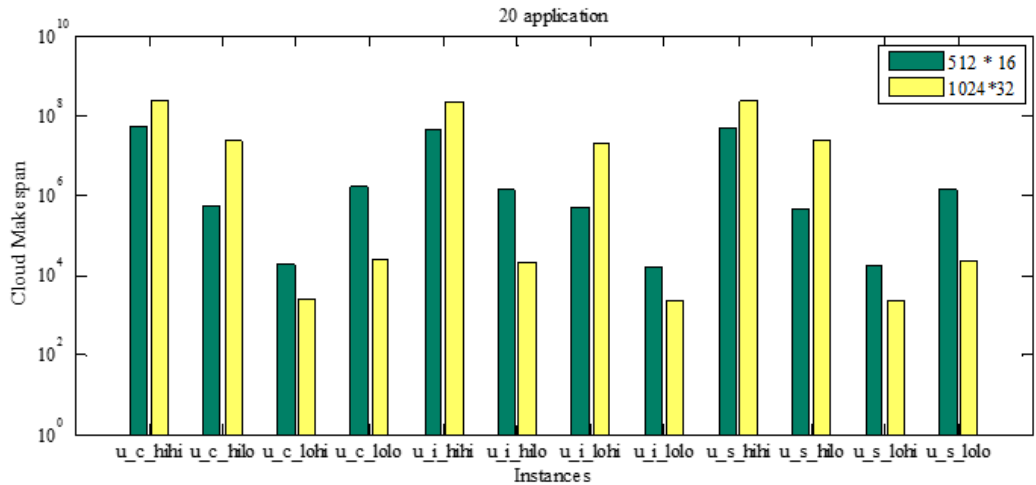

**Figure 7. Graphical comparison of makespan for 20 applications**

## 7. CONCLUSION

We have proposed a GA based algorithm to solve task scheduling problem in a multi-cloud environment. The proposed algorithm has been described with efficient encoding scheme of chromosome and fitness function. The experimental results on the proposed algorithm on two different benchmark datasets have been shown to analyze the performance of the algorithm in term of makespan. It has been shown that the algorithm performs best for the instance u_i_lolo and moderately for the instances u_c_lohi and u_s_lohi.